\documentclass[aps,prd,twocolumn]{revtex4}
\usepackage{amsmath,amssymb}

\newcommand{\eqa}{\begin{eqnarray}}
\newcommand{\neqa}{\end{eqnarray}}
\newcommand{\be}{\begin{equation}}
\newcommand{\ee}{\end{equation}}

\renewcommand{\texttt}{{}}
\usepackage{graphicx}

\begin{document}

\title{%
Fractal Structure of Loop Quantum Gravity} 

\author{Leonardo Modesto}

\affiliation{ Perimeter Institute for Theoretical Physics, 31 Caroline St., Waterloo, ON N2L 2Y5, Canada}

\date{\small\today}

\begin{abstract} \noindent
In this paper we have calculated the spectral dimension of loop quantum gravity (LQG) using simple
arguments coming from the area spectrum at different length scales. We have obtained that
the spectral dimension of the spatial section runs from $2$ to $3$, across a $1.5$ phase, 
when the energy of a probe scalar field decrees from high to low energy. We have calculated 
the spectral dimension of the space-time also using results from spin-foam models, 
obtaining a $2$-dimensional effective manifold at hight energy. Our result is consistent 
with other two approach to non perturbative quantum gravity: 
{\em causal dynamical triangulation} 
and {\em asymptotic safety quantum gravity}.

\end{abstract}

\maketitle

In the past years many approaches to quantum gravity
have studied the fractal properties of the space-time.
In particular in {\em causal dynamical triangulation} (CDT) \cite{CDT} 
and {\em asymptotic safety quantum gravity} (ASQG) \cite{ASQG} a fractal analysis 
of the space-time gives a two dimensional effective manifold
at high energy. In the two approaches the spectral
dimension is ${\mathcal D}_s=2$ at small scales and ${\mathcal D}_s=4$ al large scales.
Recently the previous ideas have been applied in the contest of {\em non commutativity} to a quantum sphere and
$\kappa$-Minkowski \cite{Dario}. In particular for the second group the author 
found a space-time spectral dimension ${\mathcal D}_s=3$ at hight energy. 
The spectral analysis is a useful tool to understand the {\em effective form}
 of the space at small and large scales. 
We think the fractal analysis could useful be also tool to predict the behavior 
of the $2$-points and $n$-points functions at small scales and to attack the 
singularity problems of general relativity in a full theory
of quantum gravity. 

 In this paper we apply to {\em loop quantum gravity}
(LQG) \cite{book1} \cite{book2} the analysis 
developed in the contest of ASQG 
by O. Lauscher and M. Reuter \cite{ASQG}.
In the contest of LQG we consider the spatial section which is
a $3d$ manifold and we extract the energy scaling of the 
metric from the area spectrum. We consider 
the $SU(2)$ representations, which appear in the area spectrum, as
continuum variables. This is a strong approximation but the result
will be consistent with the well-knowen interpretation given in \cite{book1}.
We apply the same analysis to the space-time using the area spectrum 
that is suggested by the spin-foam models \cite{DO}. 
In the space-time case the result will be consistent with the spectral 
dimension calculated in the different approach to {\em non-perturbative
quantum gravity} \cite{CDT}, \cite{ASQG}.

The paper is organizer as follow. In the first section we extract information 
about the scaling property of the $3d$ spatial section metric from the area spectrum of LQG.
The same analysis in the context of spin-foam models gives the scaling properties
of the metric in $4d$.
In second section we give a short review of the spectral dimension in diffusion 
processes. In the third section we calculate explicitly the spectral dimension of the 
spatial section in LQG and the space-time dimension using the area spectrum from
spin-foam models \cite{DO}. 
\paragraph{Scaling of the metric}
One of the strongest results of LQG is the quantization of
the area, volume and recently length operators \cite{LoopOld}.
In this section we recall the area spectrum and we deduce 
from that the energy scaling of the $3d$-metric of the spatial section.
The area spectrum on a spin-network state, 
$|\gamma; j_e, \iota_n \rangle$, without edges and nodes on the surface ${\mathcal S}$ we are considering is 
\begin{eqnarray}
\hat{A}_{\mathcal S} |\gamma; j_e, \iota_n \rangle = 
8 \pi \gamma G_N \hbar \sum_{p \bigcap S} \sqrt{j_p(j_p+1)} |\gamma; j_e, \iota_n \rangle,
\label{area}
\end{eqnarray}
where $j_p$ are the representations on the edges that cross the surface ${\mathcal S}$.
We will restrict to the case when a single edge crosses the surface.
Using (\ref{area}) we can calculate the relation between the area operator average 
for two different states of two different $SU(2)$ representations, $j$ and $j_0$, 
\begin{eqnarray}
\langle \gamma; j| \hat{A} |\gamma; j \rangle = \frac{l_P^2  \sqrt{j(j+1)}}{ l_P^2  \sqrt{j_0(j_0+1)}} 
\langle \gamma_0; j_0| \hat{A} |\gamma_0; j_0 \rangle.
\label{areass}
\end{eqnarray}
We can introduce the length square defined by $l^2=l_P^2 j$ and the infrared length 
square $l^2_0=l_P^2 j_0$.
Using this definition we obtain the scaling properties of the area's eigenvalues.
If $\langle  \hat{A}_l\rangle$ is the area average at the scale $l$ and 
$\langle  \hat{A}_{l_0}\rangle$ is the area average at the scale $l_0$ 
(with $l \leqslant l_0$) then we obtain the scaling relation 
\begin{eqnarray}
\langle  \hat{A}_l\rangle = \frac{ \sqrt{l^2(l^2+l_P^2)}}{\sqrt{l_0^2(l^2_0+l_P^2)}} 
\langle  \hat{A}_{l_0} \rangle.
\label{areall}
\end{eqnarray}

The classical area operator can be related to the spatial metric $g_{ab}$ 
in the following way.
The classical area operator can be expressed in terms of the density triad 
operator,
\begin{eqnarray}
A_{\mathcal S}=\int_{\mathcal S} \sqrt{n_a E^{a}_i n_b E^{b}_{i}} d^2\sigma,
\label{areacl}
\end{eqnarray}
and the density triad is related to the three dimensional triad by 
$e^{a}_i = E^{a}_i /\sqrt{{\rm det} E}$ and $\sqrt{{\rm det} E} = {\rm det} e$.
If we rescale the area operator by a factor ${\mathcal Q}^2$, 
$A \rightarrow A^{\prime} = {\mathcal Q}^2 A$, consequently 
the density triad scales by the same quantities, 
$E^a_i \rightarrow E^{a \prime}_i = {\mathcal Q}^2 E^a_i$.
The triad instead, using the above relation, scales as
$e^a_i \rightarrow e^{a \prime}_i = {\mathcal Q}^{-1} e^a_i$ 
and the inverse  
$e^i_a \rightarrow e^{i \prime}_a = {\mathcal Q} e^i_a$. The 
metric on the spatial section is related to the triad by 
$g_{ab} = e^i_a e^j_b \delta_{ij}$ and then it scales 
as $g_{ab} \rightarrow g_{ab}^{\prime} = {\mathcal Q}^2 g_{ab}$,
or in other words the metric scales as the area operator.

Using (\ref{areall}) and (\ref{areacl}) we obtain the following scaling for the metric 
\begin{eqnarray}
\langle  \hat{g}_{ab} \rangle_l = \Bigg[\frac{ l^2(l^2+l_P^2)}{ l_0^2(l^2_0+l_P^2)} \Bigg]^{\frac{1}{2}}
\langle  \hat{g}_{ab}\rangle_{l_0}.
\label{metricll}
\end{eqnarray}

If we want to observe the spatial section with a microscope of resolution $l$ 
we must use a probe field of momentum $k \sim 1/l$. 
The scaling property of the metric in terms of $k$ can be obtained 
replacing: $l\sim 1/k$, $l_0\sim 1/k_0$ and $l_P\sim 1/E_P$. 
Where $k_0$ is an infrared energy cutoff and $E_P$ is the Planck energy.
The scaling
of the metric in the momentum space is
\begin{eqnarray}
\langle  \hat{g}_{ab} \rangle_k= \left[\frac{ k_0^4(k^2+E_P^2)}{ k^4(k_0^2+E_P^2)} \right]^{\frac{1}{2}}
\langle  \hat{g}_{ab}\rangle_{k_0}.
\label{metrickk}
\end{eqnarray}
In particular 
we will use the scaling properties of the inverse metric,
\begin{eqnarray}
\langle  \hat{g}^{ab} \rangle_k= \left[\frac{ k^4(k_0^2+E_P^2)}{ k^4_0(k^2+E_P^2)} \right]^{\frac{1}{2}}
\langle  \hat{g}^{ab}\rangle_{k_0}.
\label{metrickkINV}
\end{eqnarray}
We define the scaling factor in (\ref{metrickkINV}), introducing a function $F(k)$: $\langle  \hat{g}^{ab} \rangle_k=F(k)\langle  \hat{g}^{ab}\rangle_{k_0}$. 
From the explicit form of $F(k)$ we have three different phases where the behavior of $F(k)$ can be approximated as follows,
\begin{eqnarray}
&& F(k)\sim1 \,\,, \,\, k \sim k_0, \nonumber \\
&& F(k) \sim k^2 \,\,, \,\,  k_0 \ll k \ll E_P, \nonumber \\
&& F(k) \sim k \,\, , \,\, k\gg E_P.
\label{Flimits}
\end{eqnarray}
We consider $F(k)$ constant for $k\lesssim k_0$; in particular we require that $F(k)\sim 1$, 
$\forall k \lesssim k_0$. To simplify the calculations without modifying the scaling
properties of the metric we introduce the new function ${\mathcal F}(k) = F(k) +1$. 
The behavior of ${\mathcal F}$ is exactly the same of (\ref{Flimits}) but with better
properties in the infrared limit useful in the calculations. 
We define here the scale function ${\mathcal F}(k)$ for future reference in the next sections,
\begin{eqnarray}
{\mathcal F}(k)= \left[\frac{ k^4(k_0^2+E_P^2)}{ k^4_0 (k^2+E_P^2)} \right]^{\frac{1}{2}} +1.
\label{Fprime}
\end{eqnarray}
   
  We can repeat the scaling analysis of this section in the case of a four dimensional
  spin-foam model. In this case the area eigenvalues are $A_j=l_P^2 j$ \cite{DO}
  and the scaling of the $4d$ metric is 
  \begin{eqnarray}
\langle  \hat{g}^{\mu \nu} \rangle_k= \frac{ k^2}{ k^2_0}
\langle  \hat{g}^{\mu \nu}\rangle_{k_0},
\label{metrickkINV4d}
\end{eqnarray}
where $\mu, \nu =1, \dots,4$. Given the explicit form of the scaling 
in (\ref{metrickkINV4d}) 
we introduce the scaling function ${\mathbb F}(k) = k^2/k_0^2 +1$.
The infrared modification introduced by hand 
does not change the hight energy behavior of the scaling function
and we can take $k \in [0, +\infty[$ in the calculations.
A different ordering in the area operator quantization can give 
a different spectrum $A_j=l_P^2  (2j+1)$ \cite{DO}, \cite{DO2}. The scaling function 
in this case is ${\mathbb G} = (k^2 (k_0^2+ 2 E_p^2))/(k_0^2 (k^2+2 E_p^2)) +1$.
Where we have introduced the usual infrared modification: $+1$.

\paragraph{The spectral dimension in diffusion processes}
We recall here the definition of spectral dimension of diffusions processes.
Consider the diffusion of a scalar probe particle on 
a $d$-dimensional Euclidean manifold with a fixed smooth metric $g_{a b}(x)$.
The heat-kernel $K_g(x, x^{\prime};T)$ gives the probability for the scalar 
test particle to diffuse from the point $x^{\prime}$ to $x$ in a diffusion time $T$.
The heat-kernel satisfies the following heat-equation 
\begin{eqnarray}
\partial_{T} K_g(x, x^{\prime};T) = \Delta_g K_g(x, x^{\prime};T),
\label{eatK}
\end{eqnarray}
where $\Delta_g = g^{-1/2} \partial_{a} ( g^{-1/2} g^{a b} \partial_{b} \phi)$
is the scalar field Laplacian.
The heat-kernel is a matrix element of the operator $\exp(T \Delta_g)$, 
$K_g(x, x^{\prime};T) =\langle x^{\prime} | \exp(T \Delta_g) | x \rangle$, 
as we can verify, 
\begin{eqnarray}
&& \partial_T K_g(x, x^{\prime};T) = \partial_T \langle x^{\prime} | {\rm e}^{T \Delta_g} | x \rangle =\nonumber \\
&& \hspace{-0.7cm} 
= \int d y \langle x^{\prime} |  \Delta_g  | y \rangle \langle y | {\rm e}^{T \Delta_g} | x \rangle =
\Delta_g K_g(x, x^{\prime};T).
\label{Kmatrix}
\end{eqnarray}
In the random walk picture the trace per unit volume, 
$P_g (T) =
V_d^{-1} \hspace{-0.2cm }\int d^d x \sqrt{g(x)} K_g(x, x;T) = V_d^{-1}{\rm Tr} \, {\rm e}^{T \Delta_g}$, 
is interpreted as an average return probability. For $T\rightarrow 0$, 
$P_g(T)$ has an asymptotic expansion, 
$P_g(T) = (4 \pi T)^{-d/2} \, \sum_{n=0}^{+\infty} c_n T^n(4 \pi T)^{d/2}$,
and for an infinitely flat space $P_g(T) = 1/(4 \pi T)^{d/2}$.
From $P_g(T)$ we can extract the dimension of the target manifold :
$d=-2 \, \partial \ln P_g(T)/\partial \ln T$.

In quantum geometry and quantum space-time we define the spectral dimension 
for a $d$-dimensional manifold 
in analogy with the classical formula, 
\begin{eqnarray}
{\mathcal D}_s = - 2 \frac{\partial \ln P_g (T)}{\partial \ln T}.
\label{spectral}
\end{eqnarray}
The formula (\ref{spectral}) will be our definition of spectral ({\em fractal}) dimension in both the spatial 
section and the space-time.

\paragraph{Spectral dimension in Quantum Gravity.}
In this section we calculate the spectral dimension of the 
spatial section in LQG and of the space-time for spin-foam models \cite{DO}. 

{\em $3d$ Spatial Section}. 
We suppose to have a smooth Riemannian metric at any energy scale $k$ that we 
denote $\langle g_{ab} \rangle_k$ and we go to probe the space at any scale $0 \lesssim k < +\infty$.
As explained in the previous section we must study the properties of the
Laplacian operator of a $3d$ manifold.
Given the scaling properties of the inverse metric (\ref{Fprime}) 
we can deduce the scaling properties of the Laplacian,
\begin{eqnarray}
\Delta(k)= {\mathcal F}(k) \Delta(k_0).
\label{laplacianscale}
\end{eqnarray}
We suppose that the diffusion process involves only a small interval of scales where ${\mathcal F}(k)$
does not change much. Under this assumption the heat-equation must contain 
$\Delta(k)$ for the specific fixed value of $k$,
\begin{eqnarray}
\partial_{T} K_g(x, x^{\prime};T) = \Delta(k) K_g(x, x^{\prime};T).
\label{heatKk}
\end{eqnarray}
We denote the eigenvalues of $\Delta(k_0)$ by $- E_n$ and introducing a resolution of
the identity in terms of the eigenvectors of $\Delta(k_0)$, we find the following solution of 
equation (\ref{heatKk}),
\begin{eqnarray}
K(x, x^{\prime}; T) = \sum_n \phi_n(x) \phi_n(x^{\prime}) {\rm e}^{-T {\mathcal F}(k) E_n}
\label{Kenergy}
\end{eqnarray}
If $\Delta(k_0)$ corresponds to the flat space 
the eigenfunctions are plane waves, $\phi_n \rightarrow \phi_p \sim \exp(i p x)$, and 
the eigenvalues of $\Delta(k_0)$ are $-E_n = - p^2$.
The eigenfunctions resolve length scales $l \sim 1/p \sim 1/\sqrt{E_n}$.
This suggest that when the manifold is probed with mode of eigenvalue $E_n$ it feels 
the metric $\langle g_{ab} \rangle_k$ for the scale $k = \sqrt{E_n}$. 
We can calculate the trace of $K(x, x^{\prime};T)$ on the plane wave basis,
\begin{eqnarray}
&& P(T) = V_d^{-1} \int d^d x \langle x| {\rm e}^{T \Delta(k)} | x \rangle \nonumber \\
&& = V_d^{-1} \int d^d x \langle x| {\rm e}^{T {\mathcal F}(k) \Delta(k_0)} | x \rangle 
\nonumber \\
&& =  \int \frac{d^d p}{(2 \pi)^d}  \, 
 {\rm e}^{- T   {\mathcal F}(k) p^2}.
\label{PTWP0}
\end{eqnarray}
Where we have introduced the spectrum of the operator $\Delta(k_0)$ and the scaling  (\ref{laplacianscale}). 
 We have said that a mode of eigenvalue $E_n$ sees a metric  $\langle g_{ab} \rangle_k$
 for the scale $k = \sqrt{E_n} = p$, then we can identify $p \equiv k$ in (\ref{PTWP0})
 obtaining 
 \begin{eqnarray}
P(T) =  \int \frac{d^d k}{(2 \pi)^d}  \, 
 {\rm e}^{- T   {\mathcal F}(k) k^2}.
\label{PTWP}
\end{eqnarray}

We have now all the ingredients to calculate the spectral dimension in LQG.
In LQG we will restrict all the previous 
integrals to the case $d=3$.

\begin{figure}
 \begin{center}
  \includegraphics[height=4cm]{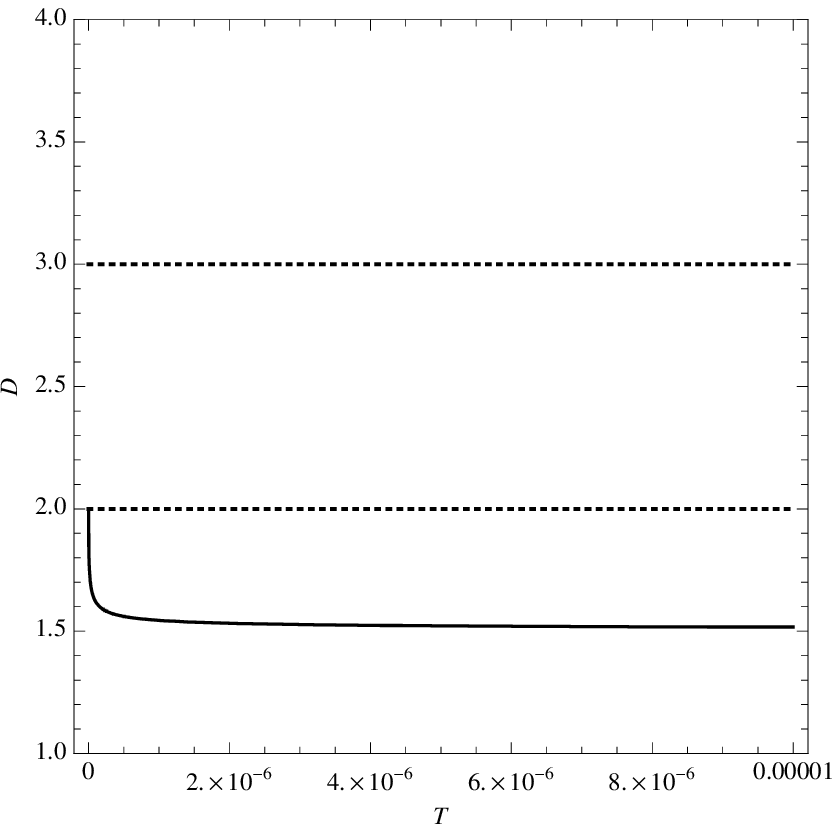}
  \hspace{0cm}
  \includegraphics[height=4cm]{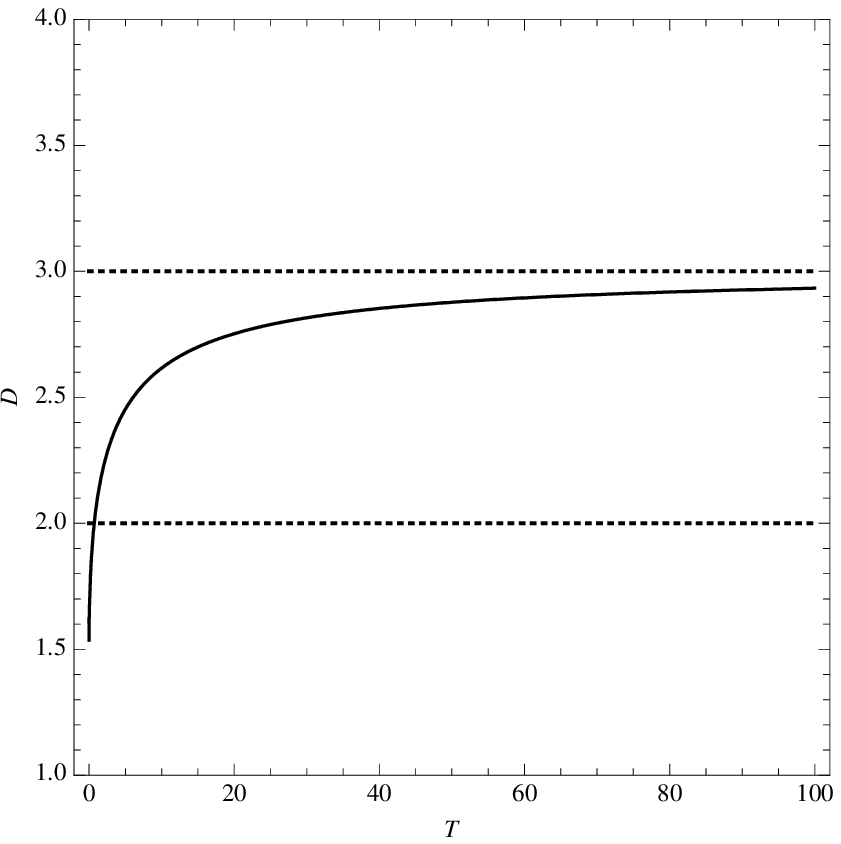}
    \end{center}
  \caption{\label{Plot1} 
 Plot of the spectral dimension as function of the diffusion time $T$.
 We can see three different phase from the left to the right.
 The first plot is 
 }
  \end{figure}

Using the relation (\ref{PTWP}) and the definition of spectral dimension (\ref{spectral})
we have 
\begin{eqnarray}
{\mathcal D}_s = 2 \, T \frac{\int d^3 k \, {\rm e}^{- k^2 {\mathcal F}(k) T} \, k^2\, {\mathcal F}(k)}{\int d^3 k \, {\rm e}^{- k^2 {\mathcal F}(k) T} }.
\label{DSLQG}
\end{eqnarray}
Given the explicit form of the scaling function ${\mathcal F}(k)$ we are not able to calculate an
analytical solution. We have calculated the spectral dimension (\ref{DSLQG}) 
numerically obtaining a function of $T$ which is plotted in Fig.\ref{Plot1}.
From the plot in Fig.\ref{Plot1}, but also calculating analytically the spectral dimension
in the three different regimes of (\ref{Flimits}), we have 
\begin{eqnarray}
{\mathcal D}_s = \left\{ \begin{array}{lll} 
         2 & {\rm for} \,\,\, k \gg E_P, \\ 
         1.5 & {\rm for} \,\,\, k_0 \ll k \ll E_P,\\
        3 & {\rm for} \,\,\, k\gtrsim k_0.
        \end{array} \right. 
\label{regimes}
\end{eqnarray}
We can conclude that in LQG we have three different phase 
that we will try to interpret in the discussion section. 

\begin{figure}
\vspace{0.15cm}
 \begin{center}
 \includegraphics[height=4cm]{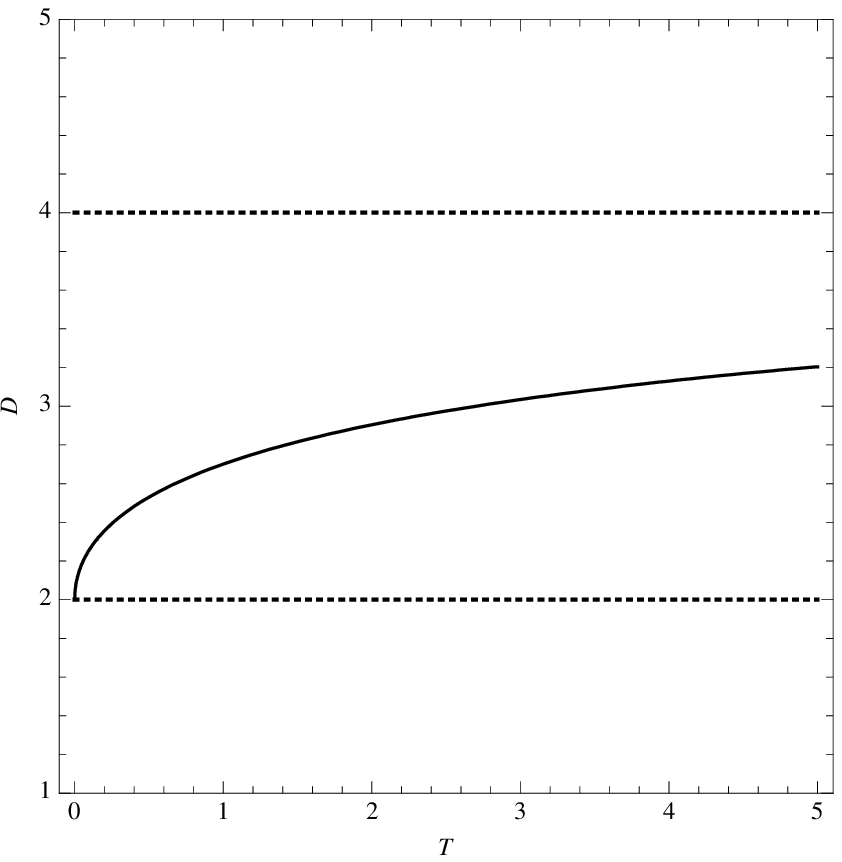}
 \includegraphics[height=4cm]{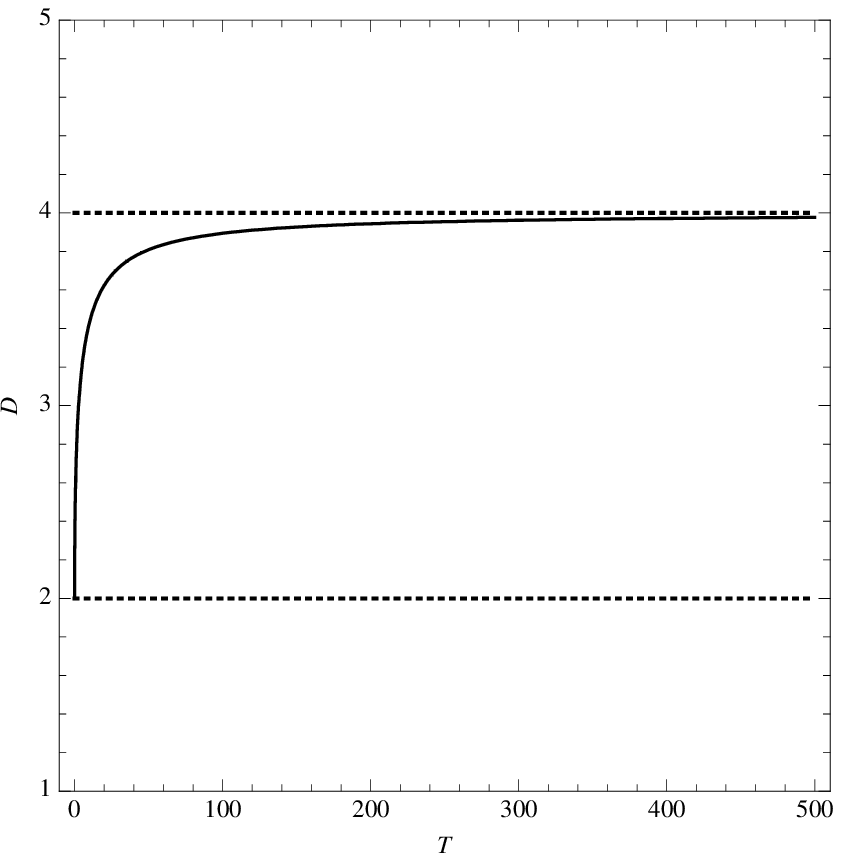}
    \end{center}
  \caption{\label{Plot4} 
Plot of the space-time spectral dimension ${\mathbb D}_s$. We have 
an hight energy phase of spectral dimension  ${\mathbb D}_s=2$ and 
a the $4d$ low energy dimension. 
 }
  \end{figure}

{\em $4d$ Space-Time.}
Using the scaling property of the space-time metric
in (\ref{metrickkINV4d}) we can calculate the
spectral dimension of the $4d$-manifold.
We use the notation $\mathbb{D}_s$ for the space-time 
spectral dimension,
\begin{eqnarray}
 {\mathbb D}_s = 2 \, T \frac{\int d^4 k \, {\rm e}^{- k^2 {\mathbb F}(k) T} \, k^2\, {\mathbb F}(k)}{\int d^4 k \, {\rm e}^{- k^2 {\mathbb F}(k) T} },
 \label{DSLQG4}
 \end{eqnarray}
 where ${\mathbb F}(k)$ is the scaling of the metric given in 
 (\ref{metrickkINV4d}). 
In Fig.\ref{Plot4} is given a plot of the spectral dimension as function 
 of the diffusion time $T$. For $T\rightarrow 0$ (or $k\sim E_p$) the we 
 obtain spectral dimension ${\mathbb D}_s =2$ and for $T \rightarrow \infty$ (or $k \rightarrow 0$)
 we obtain ${\mathbb D}_s =2$.
 We can consider the high and low energy limit obtaining the following
 behavior of the spectral dimension,
 \begin{eqnarray}
{\mathbb D}_s = \left\{ \begin{array}{lll} 
         2 & {\rm for} \,\,\, k  \gtrsim E_P, \\ 
                 4 & {\rm for} \,\,\, k \ll E_P.
        \end{array} \right. 
\label{DSLQG4dlimit}
\end{eqnarray}
Our result in space-time is in perfect accord 
 with the results in CDT \& ASQG \cite{CDT}, \cite{ASQG}.
 If we use the scaling function ${\mathbb G}(k)$ defined at the end of the section {\em a.} one
 we obtain the same behavior of the spectral dimension in the case we
 consider the ultraviolet cutoff $k < E_P$
 (this cutoff is suggested by the area spectrum $A_j = l_P^2 (2 j +1)$
 which contains the $+1$ gap on the area eigenvalue for $j=0$).
 If we consider the possibility the momentum $k \geqslant E_P$, 
 we obtain the spectral dimension ${\mathbb D}_s = 4$ for $T\rightarrow 0$ 
 (or $k\rightarrow + \infty)$. The behavior of ${\mathbb D}_s$ is instead the same
of (\ref{DSLQG4dlimit}) for $k < E_P$.
This high energy behavior of the spectral dimension is interesting if
we consider the space-time Ricci invariant ${\rm R}(g) = {\rm R}^{\mu}_{\mu}(g)$.  
Under the rescaling ${\mathbb G}(k)$ the Ricci curvature scales as:
${\rm R}(g)_k = {\mathbb G}(k) {\rm R}(g)_{k_0}$. At short distances or $k\rightarrow +\infty$
${\rm R}(g)_k$ is upper bounded as it is manifest considering the limit: 
$\lim_{k\rightarrow \infty} {\mathbb G}(k) \sim (E_p/k_0)^2$.
The upper bound of the curvature could be a sign of singularity
problem resolution showed in cosmology and black holes in the 
minisuperspace simplification of quantum gravity \cite{CBH}.
We conclude the section considering the case the area spectrum is
$A_j= l_P^2 \sqrt{j(j+1)}$. In this case the scaling function is 
the same given in (\ref{Fprime}) but the momentum $k$ is now four dimensional.
The spectral dimension has the same behavior plotted in
Fig.\ref{Plot4} in the case $k<E_P$ and instead ${\mathbb D}_s=8/3$ in
the trans-planckian limit ($k\gg E_P$). 
However if we do not consider the trans-Planckian limit 
we obtain the same spectral dimension (\ref{DSLQG4dlimit})
for any form of the area spectrum considered in this section.
\begin{figure}
 \begin{center}
 \includegraphics[height=5cm]{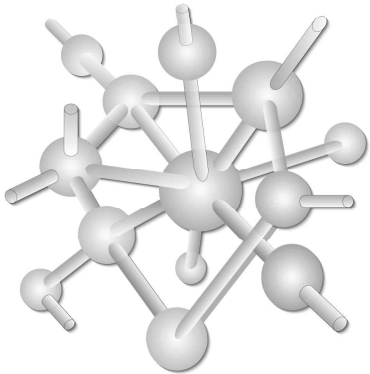}
    \end{center}
    \vspace{-0.3cm}
  \caption{\label{Plot2} 
  This artistic picture represents the possible interpretation 
  of the spectral dimension given in the discussion section of the 
  paper. The tiny connections in the picture have a regular distribution 
  in the $3d$ manifold but the $1.5$ dimension obtained in the
  paper suggest a fractal phase around the Planck scale.
  The pictures focuses on the regime when $\mathcal{D}_s \lesssim 2$.
 }
  \end{figure}

{\em Conclusions and Discussion.} 
In this paper we have calculated explicitly the spectral dimension (${\mathcal D}_s$)
of the spatial section in LQG using the area spectrum scaling
and also some strong hipotesys. We have obtained ${\mathcal D}_s$ as function
of the diffusion time $T$ or equivalently as function of the length scale.
We have three phases: a short scale phase $l \ll l_P$ of spectral 
dimension ${\mathcal D}_s = 2$, an intermediate scale phase $l_P \ll l \ll l_0$ of spectral 
dimension ${\mathcal D}_s = 1.5$ and a large scale phase of ${\mathcal D}_s = 3$.
We have calculated the spectral dimension for the space-time in
the contest of spin-foam models and we have obtained ${\mathbb D}_s =2$
at the Planck scale and ${\mathbb D}_s =4$ at low energy.
This result is same obtained in CDT \& ASQG \cite{CDT}, \cite{ASQG}.
A different area spectrum that come from a different quantum ordering 
\cite{DO} gives the same result until the Planck
scale but a new a different behavior in the trans-Planckian regime. 

We give now a possible interpretation of the spectral dimension for
the spatial section.
We can interpret the running of the spectral dimension in the following way.
First of all we want to underline that the probe scalar field is just 
a fictitious field and not a physical scalar field.
Consider a scalar field of weave length $ \lambda \ll l_P$, at this energy 
the probe field feels a $2d$ manifold, or in other words it feels the boundary Planck area
of a three dimensional chunk of space, that we can imagine of zero volume. 
The field propagates on a the $2d$ boundary of the chunk 
 of space 
until it feels tiny wormholes that connects different chunks of space. In this 
phase (we are in the regime $\lambda \gtrsim l_P$) 
the field propagates 
across the wormholes 
feeling an effective $1.5$-dimensional manifold. The initial chunk is connected 
to many others chunks then when $\lambda \gg l_P$
it feels a $3d$ manifold. 
We have assumed the connections of neighbouring  chunks
are standard wormholes but instead, because the spectral dimension $1.5$, 
the connections define a fractal structure.
In other words we can say that {\em the probe scalar field} 
({\em for} $\lambda \lesssim l_P$) 
{\em feels a fractal multi-connected manifold of ${\mathcal D}_s \lesssim 2$}.
The pictures in Fig.(\ref{Plot2}) represents {\em artistically} our interpretation. 
If we think in terms of the dual Hilbert space of spin-networks states 
the wormholes can be interpret as the two dimensional blowing up 
of the edges in the spin-network graph.

Our result is consistent with the LQG interpretation \cite{book1}. 
In LQG, on any spatial section, there is a minimum non zero area eigenvalue that
we can consider (in our interpretation)
the manifold where effectively the fictitious scalar field diffuses at high
energy. 


{\em Acknowledgements.}\\
We are grateful to  Dario Benedetti, Daniele Oriti, Carlo Rovelli, Michele Arzano and Eugenio Bianchi
for many important and clarifying discussions.

\end{document}